\documentclass[12pt]{article}
\pdfoutput=1
\usepackage{putex}

\usepackage{graphicx}
\usepackage{epstopdf}
\usepackage{enumerate}
\usepackage{cite}
\usepackage{tensor}
\usepackage{slashed}

\usepackage{hyperref}

\numberwithin{equation}{section}

\newcommand {\be} {\begin {equation}}
\newcommand {\ee} {\end {equation}}

\newcommand {\bes} {\begin {equation*}}
\newcommand {\ees} {\end {equation*}}

\newcommand{\es}[2] {\begin{equation} \label{#1} \begin{split} #2 \end{split} \end{equation}}

\newcommand{\R}{\mathbb{R}}

\newcommand{\beq}{\begin{equation}}
\newcommand{\eeq}{\end{equation}}

\begin{document}

\preprint{PUPT-2411\\MIT-CTP-4357}

\institution{IAS}{School of Natural Sciences, Institute for Advanced Study, Princeton, NJ 08540}
\institution{PU}{Department of Physics, Princeton University, Princeton, NJ 08544}
\institution{MIT}{Center for Theoretical Physics, Massachusetts Institute of Technology, Cambridge, MA 02139}

\authors{Igor R.~Klebanov,\worksat{\IAS}\footnote{On leave from
Department of Physics and Center for Theoretical Science, Princeton University.}
Tatsuma Nishioka,\worksat{\PU}  Silviu S.~Pufu,\worksat{\MIT}
Benjamin R.~Safdi\worksat{\PU}}

\date{May 2012}

\title{On Shape Dependence and RG Flow of Entanglement Entropy}

\abstract{ We use a mix of field theoretic and holographic techniques to elucidate various properties of quantum entanglement entropy.
In $(3+1)$-dimensional conformal field theory we study the divergent terms in the entropy when the entangling surface has a conical or a wedge singularity. In $(2+1)$-dimensional field theory with a mass gap we calculate, for an arbitrary smooth entanglement contour, the expansion of the entropy in inverse odd powers of the mass. We show that the shape-dependent coefficients that arise are even powers of the extrinsic curvature
and its derivatives. A useful dual construction of a $(2+1)$-dimensional theory, which allows us to exhibit these properties,
 is provided by the CGLP background. This smooth warped throat solution of 11-dimensional supergravity describes
 renormalization group flow from a conformal field theory in the UV to a gapped one in the IR. For this flow we calculate the recently introduced renormalized entanglement entropy and confirm that it is a monotonic function.
}

\maketitle

\tableofcontents

\section{Introduction}

The ground state entanglement entropies have emerged as a useful set of quantities for probing quantum entanglement and
the degrees of freedom of many-body ground states (see \cite{Eisert:2008ur,Nishioka:2009un,Calabrese:2009qy,Casini:2009sr,Solodukhin:2011gn,tadashi} for reviews and references to earlier work). If we consider the entanglement entropy (EE) of a $d$-dimensional spatial region and its complement, then
the leading term is typically proportional to the area of the $(d-1)$-dimensional boundary in units of the lattice spacing $\epsilon$. The useful information is then encoded in the sub-leading terms which depend on the shape of the boundary. For example, in $(3+1)$-dimensional CFT, it has been found
\cite{Solodukhin:2008dh} that the expansion of the entanglement entropy (EE) for a smooth closed entangling surface $\Sigma$ has the simple geometrical structure,
\es{threegen}{S= \alpha {A_\Sigma \over \epsilon^2}+
\log \epsilon \left ( \frac{a}{720 \pi} \int_\Sigma R_\Sigma + \frac{c}{240 \pi} \int_{\Sigma} (k^{\mu \nu}_a k^a_{\nu \mu} - \frac{1}{2} k_a^{\mu\mu} k^a_{\nu\nu} )\right ) \,,
 }
 where $k^a_{\mu\nu}= -\gamma_\mu^\rho \gamma_\nu^\sigma \nabla_\rho n^a_\sigma$ is the second fundamental form, $\gamma_{\mu\nu} = g_{\mu\nu} - n^a_\mu n^a_\nu$ is the induced metric (first fundamental form) on $\Sigma$, and $R_\Sigma$ is the Ricci scalar of $\Sigma$, which equals twice its Gaussian curvature.  The Weyl anomaly coefficients $a$ and $c$ are normalized above in
 such a way that $(a,c) = (n_0 + 11 n_{1/2}, n_0 + 6 n_{1/2} )$, where $n_0$ and $n_{1/2}$ are the numbers of real scalar and Dirac fields, respectively.

In $(2+1)$-dimensional CFT the structure of the entanglement entropy for a smooth contour $\Sigma$ in a plane is
\be
 S= \alpha{\ell_\Sigma \over \epsilon} - F
\ ,
\ee
where $\ell_\Sigma$ is the length of the contour.
There is no known expression for $F$ in terms of the curvature of the boundary. However, if the boundary
contains a cusp of length $r_{\text{max}}$ and opening angle $\Omega$, then $S$ contains an additional singular term
$-f_\text{cusp}(\Omega) \log (r_{\text{max}}/\epsilon)$ \cite{Casini:2006hu,Drukker:1999zq,Hirata:2006jx}.
In both field theoretic \cite{Casini:2006hu} and holographic \cite{Drukker:1999zq,Hirata:2006jx} calculations,
$f_\text{cusp}(\Omega)$ turns out to be a smooth convex function that interpolates monotonically between $\sim 1/\Omega$ behavior at small angles and zero at $\Omega=\pi$.  However, the details of the function are not universal---the holographic, free scalar, and free fermion calculations
produce different functions $f_\text{cusp}(\Omega)$.

In this paper we present new results on the shape dependence of entanglement entropy in two and three spatial dimensions, studying both
smooth and singular boundary geometries. Many of our calculations rely on the geometrical approach
to the calculation of entanglement entropy \cite{Ryu:2006bv, Ryu:2006ef, Nishioka:2006gr, Klebanov:2007ws} based on the gauge/gravity duality
\cite{Maldacena:1997re,Gubser:1998bc,Witten:1998qj}, but we also present some purely field theoretic arguments. In three spatial dimensions we will consider EE for a conical entangling surface
with opening angle $\Omega$ and
show that the calculation in $AdS_5$ produces a term $\sim \frac{\cos^2\Omega}{\sin\Omega} \log^2 (r_{\text{max}}/\epsilon)$.
For a wedge of length $L$ and opening angle $\Omega$, we will show that EE contains a divergent term $\sim f_\text{wedge}(\Omega) L/\epsilon$. Surprisingly, we find $f_\text{wedge}(\Omega)= f_\text{cusp}(\Omega)$ both in the free scalar field theory and in the holographic calculations.

In two spatial dimensions, the entanglement entropy $S(R)$ across a circle of radius $R$ has been invoked recently in deriving general constraints on renormalization group flows in $(2+1)$-dimensional field theory.
If such a field theory is conformal, then
\beq
S(R) = \alpha { 2\pi R \over \epsilon}  - F \,.
\label{genent}
\eeq
As established in \cite{Casini:2010kt, Casini:2011kv}
the subleading $R$-independent term is related to the regulated Euclidean path integral $Z$ of the CFT
on the three-dimensional sphere $S^3$: $F=-\log |Z|$. For a field theory that flows from a UV CFT to an IR CFT, it was conjectured
\cite{Myers:2010tj,Jafferis:2011zi,Casini:2011kv,Klebanov:2011gs} that $F_\text{UV} > F_\text{IR}$.
Recently, an ingenious proof of this $F$-theorem was constructed using the disk entanglement entropy \cite{Casini:2012ei}.
An important ingredient in the proof is the ``renormalized entanglement entropy'' introduced in \cite{Liu:2012ee}
\beq
{\cal F}(R)=- S(R) + R S'(R)
\ ,\label{Ffcn}\eeq
which is a finite function for theories that are conformal in the UV.
For a CFT this function takes the constant value $F$ from (\ref{genent}). An important property of ${\cal F} (R)$ is that in the limit of large $R$ it approaches $F_\text{IR}$ \cite{Liu:2012ee}. Furthermore, ${\cal F}'(R)= R S''(R)$.
It was shown in \cite{Casini:2012ei} that for any Lorentz invariant field theory $S''(R)\leq 0$. This demonstrates that
${\cal F}(R)$ is a non-increasing function and therefore proves the $F$-theorem.

In  \cite{Liu:2012ee,Ishihara:2012jg}, some holographic calculations of the renormalized entanglement entropy ${\cal F}(R)$ were presented.
Similarly, we will calculate ${\cal F}(R)$ for the smooth Cveti{\v c}-Gibbons-Lu-Pope (CGLP) solution \cite{Cvetic:2000db} of 11-dimensional supergravity,
which is a warped product of $\R^{2,1}$ and 8-dimensional Stenzel space \cite{springerlink:10.1007/BF03026543}, $\sum_{i=1}^5 z_i^2=\epsilon^2$. This smooth warped throat is similar to the KS background of type IIB string theory \cite{Klebanov:2000hb}.
The CGLP background describes RG flow from the CFT$_3$ dual to $AdS_4 \times V_{5,2}$ in the UV (for its different field theoretic descriptions, see
\cite{Martelli:2009ga,Jafferis:2009th}), to a gapped theory in the IR. The masses of some of the bound states in the CGLP theory were
calculated in \cite{Pufu:2010ie}.
Using the holographic approach to the entanglement entropy,
 we will confirm that ${\cal F}(R)$ for the CGLP background is a monotonic function that approaches zero as $\sim 1/R$ for large $R$. This function exhibits an
interesting second-order phase transition at a special value of $R$, where the bulk surface reaches the bottom of the throat and its topology changes.  Transitions of this type have been observed in earlier holographic calculations
\cite{Pakman:2008ui,Liu:2012ee} (see also \cite{Nishioka:2006gr,Klebanov:2007ws,Albash:2010mv,Albash:2012pd,Myers:2012ed}).

Generally, for theories with a mass gap of order $m$, the large $R$ expansion of the disk entanglement entropy is expected to have the form
\cite{Casini:2005zv,Casini:2009sr,Grover:2011fa,Huerta:2011qi}
\beq
S(R) = \alpha { 2\pi R \over \epsilon} +\beta\, m \, (2 \pi R) - \gamma + 2 \pi \sum_{n=0}^\infty {c_{-1-2n}\over (m R)^{2n+1}} \,,
\label{gapent}
\eeq
where $\gamma$ is the topological entangelement entropy \cite{Kitaev:2005dm,Levin:2006zz} (in the simple cases we will consider,
$\gamma$ vanishes). Following \cite{Casini:2005zv,Casini:2009sr,Huerta:2011qi} we will show that the terms
$\sim (mR)^{-2n-1}$ are related to the anomaly terms in $2n+3$ spatial dimensions.

In order to gain better insight into the structure of entanglement entropy for gapped theories, we will generalize from a circle
of radius $R$ to an arbitrary smooth contour $\Sigma$.
In this case, the general structure of the EE in gapped $(2+1)$-dimensional theories is
\es{IRgeneral}{
	S_\Sigma = \alpha \frac{\ell_{{\Sigma}}}{\epsilon}  + \beta\, m\, \ell_{\Sigma} -\gamma +  \sum_{n=0}^{\infty}
	 \frac{\tilde c_{-1 - 2n}^{\Sigma} }{m^{2n +1} }\,,
}
where the coefficients $\tilde c^\Sigma_{-1-2n}$ are integrals of functions of the extrinsic curvature and its
derivatives \cite{Grover:2011fa}.
The expansion of $S_\Sigma$ has only odd power of $1/m$ because on dimensional grounds these terms are multiplied by even powers of the extrinsic curvature and its derivatives.  Since in any pure state, and in particular in the vacuum, the EE of a region is equal to that of its complement, we have the symmetry $\kappa \to - \kappa$ \cite{Grover:2011fa}.
  Generalizing the arguments of~\cite{Casini:2005zv,Casini:2009sr,Huerta:2011qi}, we give a prescription for calculating the $\tilde c^\Sigma_{-1-2n}$ for massive free scalar and Dirac fields.  Using a holographic description of large $N$ theories with a mass gap, we calculate the coefficients $\tilde c^\Sigma_{-1}$ and  $\tilde c^\Sigma_{-3}$ explicitly.  We check the infrared expansion for the
  specific case of CGLP background.

\section{The $(2+1)$-dimensional entanglement entropy in free massive theories}  \label{massiveFree}

In this section we show how to calculate the $1/m$ expansion of the entanglement entropy for massive free scalar and fermion fields in $(2+1)$-dimensions.  We will take the entangling surface to be a smooth, closed curve $\Sigma_1$ of length $\ell_{{\Sigma_1}}$ and extrinsic curvature $\kappa$ in the $t=0$ slice of flat $\mathbb{R}^{2,1}$.

More generally, one could consider the case where the $(2+1)$-dimensional spacetime is described by a general manifold ${\cal M}$.  Using the replica trick one is then able to show that the entanglement entropy has the large mass expansion of the form given in \eqref{IRgeneral} with $\beta = -(n_0 +  n_{1/2} )/12$ (see, for example,~\cite{Fursaev:1994in,Fursaev:1995ef,Solodukhin:2011gn}).  The integers $n_0$ and $n_{1/2}$ denote the numbers of real scalar and Dirac fields, respectively, in $(2+1)$-dimensions.
The coefficients $\tilde c_{-1 - 2n}^{\Sigma_1}$ are known explicitly in the case where $\Sigma_1$ has vanishing extrinsic curvature \cite{Fursaev:1994in,Fursaev:1995ef,Solodukhin:2011gn}.
We will henceforth take ${\cal M} = \mathbb{R}^{2,1}$ and allow the surface $\Sigma_1$, which is taken to lie in the $t = 0$ plane, to have a non-trivial extrinsic curvature.  We want to determine the coefficients $\tilde c^{\Sigma_1}_{-1-2n}$ in terms of integrals of functions of the extrinsic curvature and its derivatives.  Our approach to the computation follows that of Casini and Huerta~\cite{Casini:2005zv,Casini:2009sr,Huerta:2011qi}, who showed how to compute the coefficients $\tilde c^{\Sigma_1}_{-1} = c_{-1} / R$ in the special case where $\Sigma_1$ is a circle of radius $R$.

The calculation proceeds by considering a higher, even dimensional QFT consisting of free fields in $\mathbb{R}^{2,1} \times T^k$, with $k \geq 1$ odd and $T^k$ the symmetric $k$-torus of a large volume $\Vol(T^k) = L^k$.  In the following argument one can replace $T^k$ by an arbitrary scalable $k$-dimensional smooth manifold.  We give the free fields a small mass $M$, which will act as an infrared regulator for the conformal anomaly.  We want to calculate the entanglement entropy in this theory across the $(1+k)$-dimensional surface $\Sigma_{1+k} =  \Sigma_1 \times T^k$, which fills the $k$-torus and is described by the smooth curve $\Sigma_1$ in the $t=0$ plane of $\mathbb{R}^{2,1}$.  We may Fourier decompose the field modes in the compact directions to obtain an infinite tower of massive $(2+1)$-dimensional fields, with masses
\es{massKK}{
m_{n_1, \cdots, n_k}^2= M^2+\left(\frac{2 \pi}{L} \right)^2 \sum_{i=1}^k n_i^2 \,, \qquad n_i \in \mathbb{Z} \,.
}
The entanglement entropy in the $(2+k+1)$-dimensional theory then becomes equal to the sum over $(2+1)$-dimensional entanglement entropies for massive fields across the curve $\Sigma_1$.  Taking the large $L$ limit, the spectrum of masses becomes continuous and we find
\es{Sredux}{
S_{ {\Sigma_{1+k}}}^{(2+k+1)} (M) = \frac{ k \Vol(T^k)}{2^k \pi^{k/2} \Gamma({k\over2}+1)} \int_0^{1/\epsilon} dp \, p^{k-1} S_{ {\Sigma_1}}^{(2+1)} (\sqrt{M^2 + p^2}) \,,
}
where $\epsilon$ is the UV cut-off.
We now substitute the expansion of $S_{\Sigma_1}^{(2+1)} (m)$ given in \eqref{IRgeneral} into \eqref{Sredux}.  We see that the term in the expansion of $S_{\Sigma_1}^{(2+1)} (m)$ which goes as $1/m^{k}$ determines the logarithmic conformal anomaly term in $S_{ {\Sigma_{1+k}}}^{(2+k+1)} (M) $.  Turning this argument around, suppose the entropy of the $(2 n+4)$-dimensional theory has the anomaly term
\es{anomTerm}{
\left. S_{ {\Sigma_{2 n + 2}}}^{(2 n + 4)} (M) \right|_{\text{log}} = s^{(2 n + 4)}_{\Sigma_{2 n + 2}}\log(M \epsilon) \,,
}
then  we can immediately read off the coefficient $\tilde c_{-1 - 2n}^{\Sigma_1}$:
\es{acIdent}{
\tilde c_{-1 - 2n}^{\Sigma_1} = - \frac{ \pi (2  \pi)^n (2n-1)!! }{\Vol(T^{2n+1})} s^{(2n+4)}_{\Sigma_{2 n + 2}}\,.
}
The above formula is slightly modified for fermions.  Dirac fermions in $(2n+4)$ dimensions are in a $2^{n+2}$-dimensional representation, which after dimensional reduction reduces to $2^{n+1}$ $(2+1)$-dimensional Dirac fermions.  Thus, the right hand side of \eqref{acIdent} should be divided by $2^{n+1}$ for Dirac fermions.  As a corollary to this argument, we see that the absence of the
$\log \epsilon$ terms in odd dimensional CFTs implies that the IR expansion of $S_{\Sigma_1}^{(2+1)} (m)$ contains only odd powers of $1/m$, in agreement with the arguments in \cite{Grover:2011fa}.

Let's see how this works explicitly when $n=0$.  The expression for $s_{\Sigma_2}^{(3+1)}$ is given in \eqref{threegen}.  The Euler number $ \chi(\Sigma_2) $ vanishes for $\Sigma_2 = \Sigma_1 \times S^1$.  The two normal vectors to $\Sigma_2$ are within $\mathbb{R}^{2,1}$, which we write with coordinates
\es{3dmetric}{
	ds^2_{(2+1)} &= -dt^2 + dr^2 + r^2 d\theta^2 \,,
}
where $\theta$ has period $2\pi$.
One of the normal vectors to $\Sigma_2$ is timelike, $n^1_\mu = (1,0,0,0)$, where
the fourth component is in the direction of the $S^1$ of length $L$,
and its second fundamental form vanishes.
Suppose $\Sigma_1$ is defined by a curve $r = R(\theta)$.  Then, the other normal vector is spacelike, $n^2_\mu = (0,r,-r R'(\theta),0)/\sqrt{r^2 + R'^2(\theta)}$,
and this gives a second fundamental form with non-vanishing component
\es{EC}{
	k^{2\, \theta}_\theta = \frac{R^2(\theta) + 2 R'^2(\theta) - R(\theta)R''(\theta)}{(R^2(\theta) + (R'(\theta))^2)^{3/2}} \equiv \kappa (\theta)\ ,
}
where $\kappa(\theta)$ is the extrinsic curvature of the surface $\Sigma_1$ in the $\mathbb{R}^2$ plane.
It follows that both $k^a_{\mu\nu}k_a^{\mu\nu}$ and $k_a^{\mu\mu}k^a_{\nu\nu}$ in \eqref{threegen} become $\kappa^2(\theta)$.  This leads to
\es{a2Rulekappa}{
\tilde c^{\Sigma_1}_{-1} = -\frac{1}{480} (n_0 + 3 n_{1/2} ) \oint ds\, \kappa^2 \,
}
and
\es{Sto1}{
S_{\Sigma_1}^{(2+1)} (m) = \alpha \frac{\ell_{{\Sigma_1}}}{\epsilon} - \frac{m (n_0 + n_{1/2} )  \ell_{{\Sigma1}} }{12} -\frac{ n_0 + 3 n_{1/2} }{480\,m} \oint ds\, \kappa^2 + O(1/m^3) \,,
}
where we stress that $n_{1/2}$ the number of $(2+1)$-dimensional Dirac fermions.

In principle the calculation of the higher order corrections to the entanglement entropy in powers of $1/m$ would proceed analogously.  For example, to calculate the coefficient $\tilde c_{-3}^{\Sigma_1}$, which gives the order $1/m^3$ correction to the entropy, we would need to first calculate $s^{(5+1)}_{\Sigma_{4}}$, with $\Sigma_4 = \Sigma_1 \times T^3$.  
We then expect
\es{sigma5}{
  s^{(5+1)}_{\Sigma_{4}} = \Vol(T^3) \left[(A_0 \, n_0 +A_{1/2} \, n_{1/2}^{(6)} ) \oint ds \, \kappa^4 + (B_0 \, n_0 + B_{1/2} \, n_{1/2}^{(6)}) \oint ds \, \left( { d \kappa \over ds} \right)^2 \right] \ ,
  }
  for some coefficients $(A_0, A_{1/2} )$ and $(B_0, B_{1/2})$, which should be functions of the $6$-dimensional anomaly coefficients.  We use the notation $n_{1/2}^{(6)} $ to stress that this counts the number of $(5+1)$-dimensional Dirac fermions.  This then leads to
  \es{a3}{
 \tilde c_{-3}^{\Sigma_1} =-2 \pi^2  \left[ \left(A_0 \, n_0 + \frac{A_{1/2}}{4} \, n_{1/2} \right) \oint ds \, \kappa^4 + \left(B_0 \, n_0 + \frac{B_{1/2}}{4} \, n_{1/2} \right) \oint ds \, \left( { d \kappa \over ds} \right)^2 \right] \,.
  }
  This formula is consistent with the general arguments in \cite{Grover:2011fa}.

\section{Holographic computation of the $(2+1)$-dimensional entanglement entropy in gapped backgrounds}  \label{general} \label{Holog}

The (renormalized) entanglement entropy may be calculated holographically by following the usual procedure for holographic entanglement entropy~\cite{Ryu:2006bv, Ryu:2006ef, Nishioka:2006gr, Klebanov:2007ws}.  Consider a $(d+1)$-dimensional large $N$ field theory with a $D$-dimensional gravitational dual.  While we will ultimately be interested in $(2+1)$-dimensional QFT, for now we keep the dimension $d$ general.  As in~\cite{Klebanov:2007ws}, let the gravitational background have the Einstein-frame metric
\es{ESframe}{
ds_{D}^2 =  \alpha(u) [ du^2 + \beta(u) dx^{\mu} dx_{\mu} ] + g_{ij} dy^i dy^j \,, \qquad dx^{\mu} dx_{\mu} = -dt^2 + dr^2 + r^2 d \Omega_{d-1}^2 \,,
}
with $\alpha(u) > 0$ and $\beta(u) > 0$ and $i,j = d+3, \cdots,D$.  The compact, internal $(D - d - 2)$-dimensional manifold is taken to have a volume
\es{VolU}{
V(u) \equiv \int \prod_{i = d+3}^{D} dy^i \sqrt{\det g} \,,
}
which is a function of the holographic radial coordinate $u$.  We assume that $u$ has a minimal value $u_0$ where a $p$-sphere in the internal manifold shrinks to zero size, resulting in $V(u_0)=0$.  At $u_0$ we assume that all supergravity fields are regular, which implies $\alpha(u_0)$ and $\beta(u_0)$ are finite.  The coordinate $u$ ranges from infinity in the far UV to $u_0$ in the far IR.  Such geometries typically describe confining gauge theories.

We further assume that the gravitational theory approaches a conformal fixed point in the UV ($u = \infty$), and we work in coordinates where
\es{fixedpointcond}{
\lim_{u \to \infty} \alpha(u) = \alpha_{\text{UV}}   \,, \qquad \, \lim_{u \to \infty} V(u) = V_{\text{UV}}  \,, \qquad  \beta(u) = \exp \left({{2 u \sqrt{ \alpha_{\text{UV}} }  \over L_{\text{UV}}  }} \right) + \ldots \,,
}
where $ \alpha_{\text{UV}} $ and $ V_{\text{UV}}$ are constants, and $L_{\text{UV}}$ is the radius of $AdS_{d+2}$.

We want to calculate the entanglement entropy in the QFT across a codimension two spacelike surface $\Sigma_{d-1}$.  The entanglement entropy~\cite{Ryu:2006bv, Ryu:2006ef, Nishioka:2006gr, Klebanov:2007ws} is calculated holographically by finding the $(D-2)$-dimensional surface $\Sigma_{D-2}$, which approaches $\Sigma_{d-1}$ at the boundary of the bulk manifold, is extended in the rest of the spatial dimensions, and minimizes the area functional
\es{action}{
S_\Sigma = \frac{1}{4 G_N^{(D)}} \int_{\Sigma_{D-2}} d^{D-2} \sigma \sqrt{G_{\text{ind}}^{(D-2)}} \,,
}
where $G_{\text{ind}}^{(D-2)}$ is the induced metric on $\Sigma_{D-2}$.   The entanglement entropy is then given by the functional $S_\Sigma$ evaluated at the extremum.

A case of particular interest is when the region $\Sigma_{d-1}$ is the $(d-1)$-sphere of radius $R$.  Writing the radial coordinate $r$ as a function of the holographic coordinate $u$, the induced metric on $\Sigma_{D-2}$ is
\es{dsSigGen}{
ds_{\Sigma}^2 = \alpha(u) [ (1 + \beta(u) (\partial_u r)^2 ) du^2 + \beta(u) r^2(u) d\Omega_{d-1}^2 ] + g_{ij} dy^i dy^j \,,
}
which gives the following expression for the area functional in terms of the unknown function $r(u)$:
\es{areaGen}{
S(R) &= \frac{\text{Vol}(S^{d-1})}{4 G_N^{(D)}} \int_{u_0}^\infty du \, r^{d-1}(u) g(u) \sqrt{1 + \beta(u) (\partial_u r)^2} \,, \\
g(u) &= \alpha^{d/2}(u) \beta^{(d-1)/2}(u) V(u) \,.
}
In general we need to first solve the Euler-Lagrange equation,
\es{ELGen}{
(d-1) r^{d-2}(u) g(u) \sqrt{1 + \beta(u) (\partial_u r)^2} &= \frac{d}{d u} \left[ \frac{r^{d-1}(u) g(u) \beta(u) (\partial_u r) }{\sqrt{1 + \beta(u) (\partial_u r)^2} } \right] \,,
}
for the function $r(u)$, then evaluate the area functional in \eqref{areaGen} on the solution with a UV cut-off $u < u_{\text{UV}}$, then use \eqref{Ffcn} to construct the finite renormalized entanglement entropy.  For non-trivial backgrounds this must be done numerically.
To solve the equation of motion \eqref{ELGen}, we also need to specify the boundary conditions. There are two types of solutions with different topologies.

One of them, which we will call the cylinder-type solution, terminates at $u=u_0$ where
the volume of the internal space becomes zero: $V(u_0)=0$.
One can find the form of the solutions $r(u)$ for $u$ near $u_0$ by expanding \eqref{ELGen} around $u=u_0$:
\es{CylExpand}{
r(u) &= r_0 + \frac{d-1}{4 r_0 \beta (u_0)} (u-u_0)^2 + O ((u-u_0)^3) \,, \qquad r_0 > 0 \,.
}

The other type of solution, which we call the disk-type solution, has
a tip at $u=u_{\text{min}} > u_0$, where the radius of the sphere becomes zero: $r(u_{\text{min}})=0$.
For $u$ near $u_{\text{min}}$, the solutions behave like
\es{CapExpand}{
r(u) = 2\sqrt{ \frac{d g(u_{\text{min}} )}{2\beta (u_{\text{min}}) g'(u_{\text{min}}) + g(u_{\text{min}}) \beta' (u_{\text{min}})}} (u-u_{\text{min}})^{1/2} + O( (u-u_{\text{min}})^{3/2})\,.
}

\subsection{IR behavior of the EE for a circle}

We may obtain the IR asymptotic behavior of the entanglement entropy for $\Sigma_{1} = S^{1}$ through an analytic procedure, and in doing so we show that the renormalized entanglement entropy approaches zero in the IR from above like $1/R$, where $R$ is the radius of the $S^1$.  Note that in this section we restrict to the physical dimension $d=2$.  In the following section we generalize the computation by allowing for a general entangling surface $\Sigma_{1}$.

 For now, we take $\Sigma_{1} = S^{1}$ of radius $R$.  We assume that at large $R$ the solutions to the Euler-Lagrange equations will be of the form $r(u) = R + \delta(u) / R$, with $\delta(u)$ independent of $R$.  Expanding the Euler-Lagrange equation in powers of $1/R$, we find the equation
\es{ELpert}{
\frac{d}{d u} [ g(u) \beta(u) \delta'(u) ] = g(u) \,,
}
which may be integrated to obtain
\es{asympInt}{
\delta(u) &= - \int_u^{\infty} du' {1 \over g(u') \beta(u') } \int_{u_0}^{u'} du'' \, g(u'') \,.
}
Expanding the area functional in \eqref{areaGen} and using the equation of motion in \eqref{ELpert} gives
\es{AreaPert}{
S(R) =  \frac{ 2 \pi }{4 G_N^{(D)}}
\left[ R {V_{\text{UV}} L_{\text{UV}} \over \epsilon} + R \left( \int_{u_0}^{u_\infty} du g(u) -  {V_{\text{UV}} L_{\text{UV}} \over \epsilon} \right) \right. \\
\left.  - \frac{1}{2 R}  \int_{u_0}^{\infty} du \, g(u) \beta(u) [\delta' (u)]^2  + O(R^{-3}) \right] \,,
}
where we used the boundary conditions $\delta(u_\infty) = 0$ and $\delta'(u_0)=0$ for the  solution,
which make the surface term vanish.
The UV cut-off $\epsilon$ defined by
\es{epsilonDef}{
{1 \over \epsilon^2} = \alpha_{\text{UV}} \exp\left({2 u_{\infty} \sqrt{\alpha_{\text{UV}}} \over L_{\text{UV}} } \right) \,.
}

To compare with \eqref{IRgeneral}, we set the mass $m$ to unity and use the dimensionless radius $R$ for convenience.
We then see that we can make the identifications
\es{compareIRgen}{
\alpha &= { V_{\text{UV}} L_{\text{UV}} \over 4 G_N^{(D)} } \,, \qquad \beta =  { 1 \over 4 G_N^{(D)} } \left(  \int_{u_0}^{u_\infty} du g(u) -  {V_{\text{UV}} L_{\text{UV}} \over \epsilon} \right) \,,  \\
\tilde c_{-1}^\Sigma &= {-1 \over 8 G_N^{(D)} } \int_{u_0}^\infty {d u \over g(u) \beta(u) } \left( \int_{u_0}^\infty du' g(u') \right)^2 \, \oint ds \,\kappa^2  \,.
}
Notice that the coefficient $\beta$ is finite and independent of the UV cut-off $\epsilon$.  To calculate the coefficients $\tilde c_{-3}^\Sigma$ we must consider a more general entangling surface.  This is because $d \kappa / ds = 0$ for the circle.  In the following section we generalize the above calculation to allow for a general, smooth entangling surface, and in doing so we calculate $\tilde c_{-3}^\Sigma$.

\subsection{IR behavior of the EE for a general entangling surface} \label{ss:genent}

We would like to repeat the calculation in the previous section
allowing for a general spacelike entangling surface ${\Sigma_1}$.  While we believe that the computation can be carried out in full generality, it is enough to restrict ourselves to a closed curve $\Sigma_1$ that is a boundary of a star-shaped domain.\footnote{A star-shaped domain is a set $S \subset \R^n$ with the property that there exists a point $x_0 \in S$ such that the line segments joining $x_0$ to all other points in $S$ are contained in $S$.}  Such a curve can be parameterized using polar coordinates by a function $R_\Lambda(\theta)$.  We write the entangling surface as $R_\Lambda(\theta) = \Lambda R(\theta)$, with $ R(\theta)$ a smooth function and $\Lambda \geq 1$.  The IR limit corresponds to $\Lambda$ large enough such that the extrinsic curvature is small, $\kappa_\Lambda(\theta) = \Lambda^{-1} \kappa(\theta) \ll 1$, along the entire curve.

The induced metric on the bulk surface $\Sigma_{D-2}$ is now
\es{dsSigGenRth}{
ds_{\Sigma}^2 &= \alpha(u) \left[ (1 + \beta(u) (\partial_u r)^2 ) du^2 + \beta(u) (r^2(u,\theta) + \big(\partial_\theta r)^2 \big) d\theta^2 \right. \\
& \left. + \,2 \beta(u) (\partial_\theta r)( \partial_u r)d\theta \,d u\right] + g_{ij} dy^i dy^j \,,
}
where the radial coordinate $r(u,\theta)$ is taken to be a function of the holographic coordinate $u$ and the angular coordinate $\theta$.  We require that $\lim_{u \to \infty} r(u,\theta) = R_\Lambda(\theta)$.  The area functional for the entanglement entropy may be written as
\es{areaGenRth}{
S_{\Sigma} &= \frac{1}{4 G_N^{(D)}} \int_0^{2 \pi} d \theta \,\int_{u_0}^\infty du \, g(u)  \sqrt{\left(r^2(u, \theta) + (\partial_\theta r)^2 \right) \left(1 + \beta(u) (\partial_u r)^2 \right) - \beta(u) (\partial_\theta r)^2( \partial_u r)^2} \,,
}
with $g(u)$ and $\beta(u)$ defined as before.

We assume that in the IR the solutions to the Euler-Lagrange equation give
\es{genAnsatzRth}{
r(u,\theta) = \Lambda R(\theta) + \frac{\delta(u,\theta)}{\Lambda}+ O(1/\Lambda^3) \,,
}
with $\delta(u,\theta)$ order $\Lambda^{0}$.  We Substitute the ansatz in \eqref{genAnsatzRth} into the area functional in \eqref{areaGenRth} and expand in inverse powers of $\Lambda$ up to and including terms of order $1/\Lambda$:
\es{areaGenRthPert}{
S_{\Sigma}  &= \frac{1}{4 G_N^{(D)}} \int_0^{2 \pi} d \theta \,\int_{u_0}^\infty du \,  \left[ \Lambda g(u) \sqrt{R(\theta)^2 + R'(\theta)^2 }  \right. \\
&\left. +  \frac{g(u) R(\theta)}{\Lambda} \left( \frac{R(\theta) \beta(u)}{ \sqrt{R^2(\theta) + R'(\theta)^2} } (\partial_u \delta)^2 +2 \kappa(\theta) \delta(u,\theta) \right) \right] + O(1/\Lambda^3)\,,
}
where the extrinsic curvature of the entangling surface, $\kappa(\theta)$, is given explicitly in \eqref{EC}.  Applying the variational principle to find the Euler-Lagrange equation for $\delta(u,\theta)$ gives
\es{ELpertRth}{
\frac{d}{d u} [g(u) \beta(u) \partial_u \delta(u,\theta) ] ={\kappa(\theta) \sqrt{R(\theta)^2 + R'(\theta)^2} \over R(\theta)} g(u)  \,,
}
which may be integrated to give
\es{asympIntRth}{
\delta(u,\theta) &= - { \kappa(\theta) \sqrt{R(\theta)^2 + R'(\theta)^2}  \over R(\theta)}  \int_u^{\infty} du' {1\over g(u') \beta(u') } \int_{u_0}^{u'} du'' \, g(u'') \,.
}

We want to calculate the terms in the expansion of $S_{\Sigma}$ of order $1/\Lambda^3$.  These terms are completely determined by the expansion of $r(u,\theta)$ in \eqref{genAnsatzRth} through order $1/\Lambda$.  Expanding the area function in \eqref{areaGenRth} through order $1/\Lambda^3$ and evaluating on the solution for $\delta(u,\theta)$ given in \eqref{asympIntRth} allows us to determine the $\tilde c_{-3}^\Sigma$ coefficients in \eqref{IRgeneral}:
\es{entanglePertRth3}{
\tilde c_{-3}^\Sigma =  a_{-3}^{(1)} \left( \frac12 \oint ds\, \kappa^4 - \oint ds\, \left( \frac{d \kappa}{ds} \right)^2 \right) + a_{-3}^{(2)}  \oint ds\, \kappa^4 \,,
}
with
\es{am3}{
a_{-3}^{(1)} &=  -\frac{1}{4 G_N^{(D)}}  \int_{u_0}^\infty \frac{du}{g(u) \beta(u)} \left[ \int_{u_0}^u du' g(u') \right]^2    \int_u^{\infty}  {du'\over g(u') \beta(u') } \int_{u_0}^{u'} du'' \, g(u'') \,, \\
a_{-3}^{(2)} &=  -\frac{1}{32 G_N^{(D)}} \left( \oint ds \, \kappa^4 \right) \int_{u_0}^\infty \frac{d u}{g(u)^3 \beta(u)^2} \left( \int_{u_0}^u du' \, g(u') \right)^4 \,.
}

\section{An example: CGLP background of M-theory} \label{CGLP}

 The CGLP background~\cite{Cvetic:2000db} of M-theory is the gravitational dual of a gapped $(2+1)$-dimensional field theory, which nicely illustrates the general features discussed in the previous section.
 The supergravity background is a warped product of $\mathbb{R}^{2,1}$ and an eight-dimensional Stenzel space~\cite{springerlink:10.1007/BF03026543}
 \es{subset}{
 \sum_{i=1}^5 z_i^2 = \varepsilon^2 \,,
 }
 where $\varepsilon$ is a real deformation parameter.  When $\varepsilon = 0$ this equation describes an eight-dimensional cone whose base is the Stiefel manifold $V_{5, 2}$.

 As explained in \cite{Cvetic:2000db,Klebanov:2010qs}, the Stenzel space \eqref{subset} can be parameterized by a radial coordinate $\tau$ ranging from $0$ to $\infty$ and the seven angles in $V_{5, 2}$.  At $\tau =0$ a $3$-sphere shrinks to zero size, and the $\tau = 0$ section is a round $S^4$.

 The $11$-dimensional metric  is of the form of the metric in \eqref{ESframe} if we identify the holographic coordinate $u$ with $\tau$, where $\tau_0 = 0$, and\footnote{We follow the conventions of  \cite{Klebanov:2010qs} and work in units where $\varepsilon = 1$.}
 \es{alphabetaCGLP}{
 \alpha(\tau) &= \frac{H^{1/3}(\tau) c^2(\tau)}{4} \,, \qquad \beta(\tau) = \frac{4}{c^2(\tau) H(\tau)} \,, \\
  V(\tau) &= \frac{9}{2} 3^{1/8} \pi^4 H^{7/6}(\tau) ( 2 + \cosh \tau )^{3/8} \sinh^{3/2}\left( \frac{\tau}{2} \right) \sinh^{3/2}( \tau )\,.
 }
The functions $H(\tau)$ and $c(\tau)$ are defined by
\es{Hcdeff}{
 H(\tau) = {(2\pi \ell_P)^6 N\over 81 \pi^4} 2^{3/2} 3^{11/4} \int_{(2 + \cosh \tau )^{1/4}}^\infty \frac{dt}{(t^4-1)^{5/2}} \,, \qquad  c^2(\tau) = { 3^{7/4} \over 2}  \frac{ \cosh^3 {\tau \over 2}}{ (2+ \cosh \tau)^{3/4}} \,,
 }
where $N$ is the number of units of asymptotic $G_4$ flux.  In particular, notice that $V(\tau = 0) = 0$, which is a result of the vanishing $3$-sphere.  For more details on the CGLP background see, for example,~\cite{Cvetic:2000db,Klebanov:2010qs}.

 We begin by studying the entanglement entropy and renormalized entanglement entropy in the simpler case where the entangling region $\Sigma_1$ is taken to be a circle of radius $R$.  In this case the Euler-Lagrange equation for the function $r(\tau)$ in \eqref{ELGen} may be solved numerically with the boundary condition $r(\tau = \infty) = R$.  In practice, we cut the space off at some large $\tau_{\text{UV}}$.  For each $R > 0$ there exists a value $\tau_{\text{min}} (R)$, which is the smallest value of $\tau$ for which the function $r(\tau)$ is defined.  There exists a critical value $R_{\text{crit}} \approx .73$ for which $r(\tau_{\text{min}}  = 0) = 0$.  For $R < R_{\text{crit}}$ the solutions to the equation of motion describe surfaces of disk type that behave as in \eqref{CapExpand} for $\tau$ near $\tau_{\text{min}}$.  The topology of these surfaces is that of a disk times $V_{5, 2}$. The solutions for $R > R_{\text{crit}}$ are surfaces of cylindrical type that
 stretch to the bottom of the Stenzel space and behave as in \eqref{CylExpand} for $\tau$ near $\tau_0 =0$.  The topology of these surfaces is that of a circle times the Stenzel space.

\begin{figure}[htb]
 \leavevmode
\begin{center}$
 \mbox{\bf (a)} \scalebox{.8}{\includegraphics{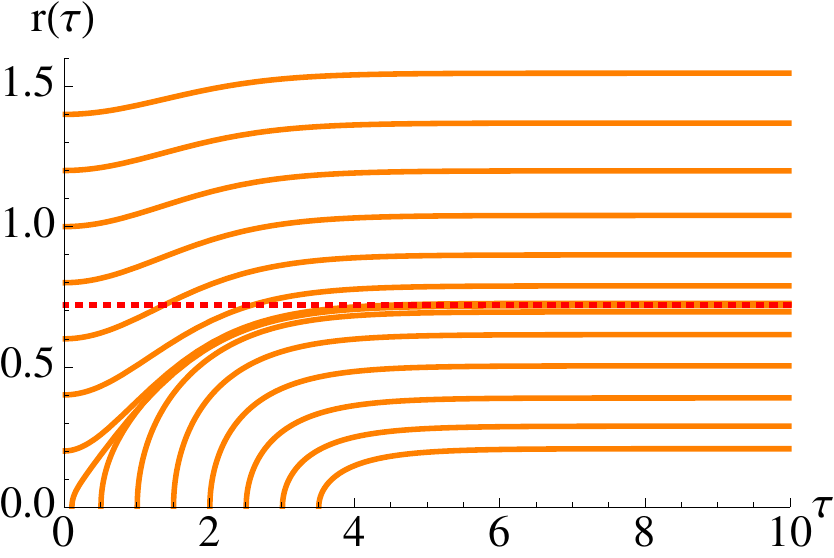}} \newline
\begin{array}{cc}
 \mbox{\bf (b)}  \scalebox{.8}{\includegraphics{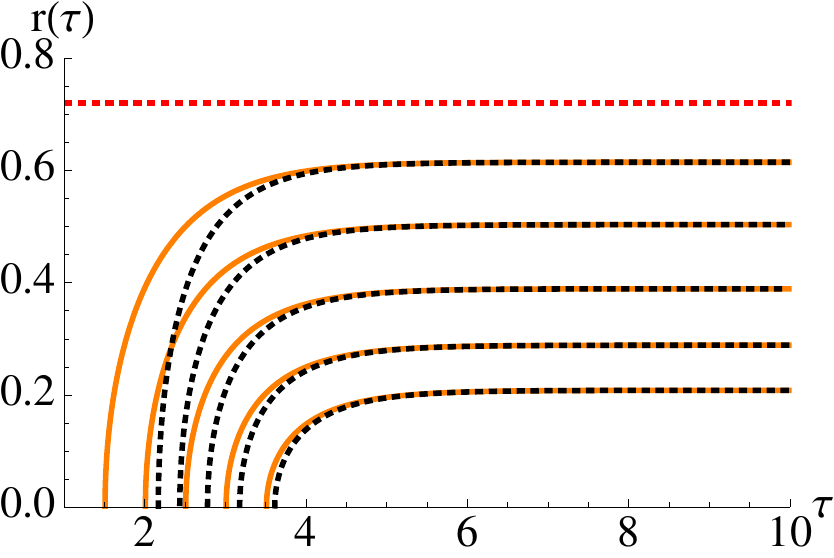}} &  \mbox{\bf (c)}  \scalebox{.8}{\includegraphics{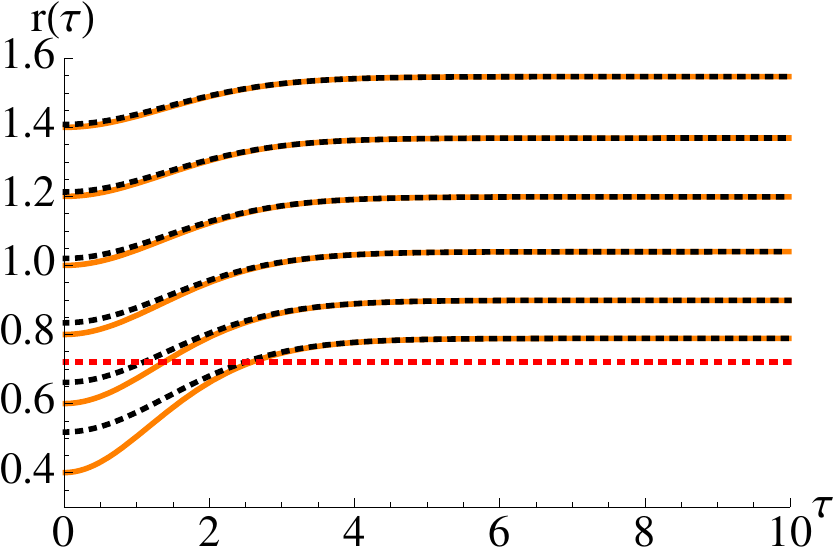}}
\end{array}$
\end{center}
\caption{{\bf (a)}  Numerical solutions to the equation of motion for the holographic entangling surface, given by $r(\tau)$, in the CGLP theory.  The dotted red line indicates the critical value $R_{\text{crit}}$, where the solutions change from disk-type to cylinder-type. {\bf (b)}  A zoomed-in plot of the UV region, with disk-type solutions, where we plot the AdS approximation in \eqref{UVfcn} in dotted black. {\bf (c)} A zoomed-in plot of the IR region, with cylinder-type solutions, with the analytic approximation given by $\delta(\tau)$ in \eqref{asympInt} plotted in dotted black.}
\label{TOTplot}
\end{figure}

 In Figure~\ref{TOTplot} (a)
 we plot the numerical solutions to the equation of motion for a range of $R < R_{\text{crit}}$ and $R > R_{\text{crit}}$.  In the far UV the solution for $r(\tau)$ should approach the AdS solution
 \es{UVfcn}{
 r(\tau) = \sqrt{ R^2 - 2^{5/2} 3^{1/2} e^{-3 \tau / 2} } \,.
 }
 In Figure~\ref{TOTplot} (b) we zoom in on some of the disk-type solutions in the far UV and  plot the solutions along with the asymptotic in \eqref{UVfcn}.  In the far IR region the cylinder-type solutions should be well approximated by the function $\delta(\tau)$ in \eqref{asympInt}.  In Figure~\ref{TOTplot} (c) we plot some of the cylinder-type solutions along with the analytic approximation.

\begin{figure}[htb]
\begin{center}
\leavevmode
\scalebox{1}{\includegraphics{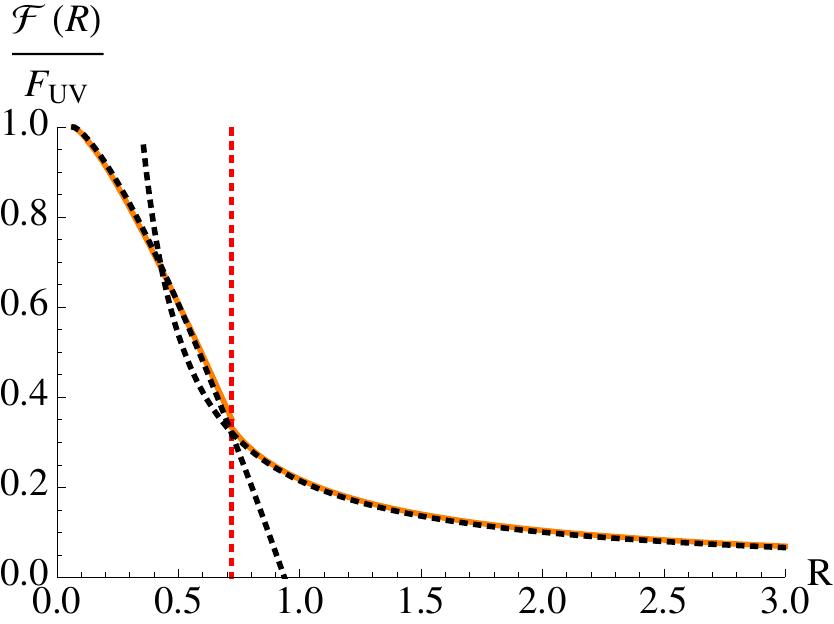}}
\end{center}
\caption{The renormalized entanglement entropy ${\cal F}(R)$ along the RG flow in the CGLP theory plotted in orange.  The left dotted black curve is the asymptotic UV approximation to ${\cal F}(R)$ given in \eqref{Fuv}.  The right dotted black curve is the IR approximation to ${\cal F}(R)$ given in \eqref{IRFCGLP}.  The dotted red line marks the value $R = R_{\text{crit}}$.}
\label{CGLPF}
\end{figure}

  As was discussed in section~\ref{general}, to calculate the renormalized entanglement entropy it is sufficient to evaluate the entanglement entropy with a strict UV cut-off.  We cut off the space at a large $\tau$ value $\tau_{\text{UV}}$.  We then numerically integrate the area functional  and differentiate it to construct ${\cal F}$.  A plot of the renormalized entanglement entropy along the RG flow is given in Figure~\ref{CGLPF}.

 \subsection{The renormalized entanglement entropy in the UV and the IR }

 In the far UV we can treat the CGLP M-theory background as a perturbation of the $AdS_4 \times V_{5,2}$ background.  From \eqref{SRconf1}, we know that the UV fixed point has a renormalized entanglement entropy
 \es{FV52}{
F_{\text{UV}} = \frac{16 \pi N^{3/2}}{27} + O(N^{1/2}) \,,
 }
where we have used $\Vol ( V_{5,2} )= 27 \pi^4 / 128$.  To describe the RG flow in the vicinity of the UV fixed point, it is convenient to use the effective $4$-dimensional metric in the form of \eqref{D4Metric2}.  A straightforward calculation shows that at small $y$ the function $f(y)$ has the expansion
\es{fyCGLP}{
f(y) = 1 + 2^{1/3} 3^{2/3}  y^{4/3} + \cdots\,,
}
which, using \eqref{fexpand}, implies the RG flow is driven by an operator in the UV field theory of dimension $\Delta = 7/3$, which is consistent with~\cite{Klebanov:2010qs}.  Using \eqref{FFirstOrder}, we then see that at small $R$
 \es{Fuv}{
 {\cal F}(R) =  \frac{16 \pi N^{3/2}}{27} \left( 1 - {3 \over 7} 2^{1/3} 3^{2/3} R^{4/3} + \cdots \right) \,.
 }
 This function is plotted together with the numerical solution in Figure~\ref{CGLPF}.  Note that it is a very good approximation to the actual renormalized entanglement entropy for $R < R_{\text{crit}}$.  We also see explicitly that $\partial_R {\cal F} = 0$ at $R = 0$.

 In the IR we may use \eqref{compareIRgen} and~\eqref{entanglePertRth3} to get an asymptotic expression for the renormalized entanglement entropy, which gives
 \es{IRFCGLP}{
 {\cal F} \approx  \frac{16 \pi N^{3/2}}{27} \left( \frac{0.1959}{R} +\frac{1.845\times 10^{-2}}{R^3}  + O(1/R^5)\right) \,.
 }
 This function is plotted in Figure~\ref{CGLPF}, which shows that it is a good approximation to the actual renormalized entanglement entropy at large $R$.

 \subsection{Tests of the shape dependence of the entanglement entropy}

 In this section we will consider a more general spacelike entangling surface $\Sigma_1$, specified by the function $R(\theta)$ in polar coordinates.  We want to check \eqref{asympIntRth}, which gives an approximation to the cylinder-type solutions in the far IR.  In particular, this equation claims that the variation of the bulk entangling surface $\Sigma_2$ away from the straight cylinder is proportional to the combination $\kappa(\theta) ( \sqrt{R(\theta)^2 + R'(\theta)^2 } / R(\theta) )$.

\begin{figure}[htb]
 \leavevmode
\begin{center}$
\begin{array}{cc}
 \mbox{\bf (a)}  \scalebox{.8}{\includegraphics{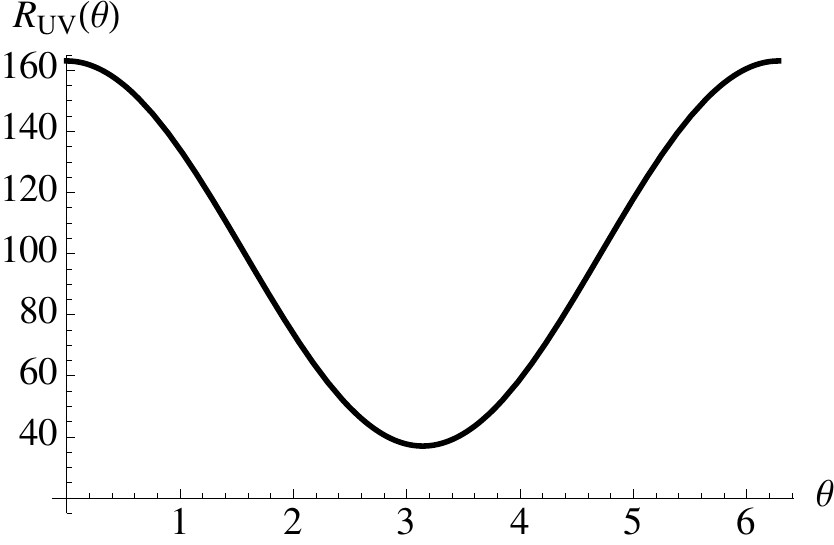}} &  \mbox{\bf (b)}  \scalebox{.88}{\includegraphics{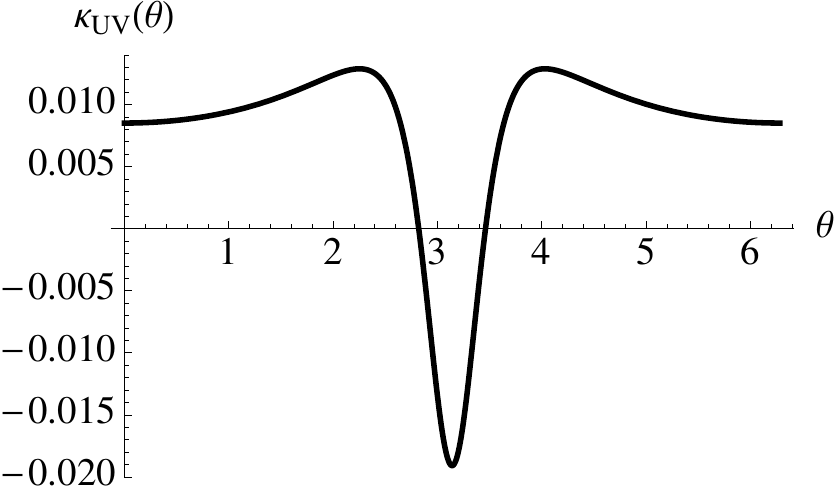}}
\end{array}$
\end{center}
\caption{{\bf (a)} A plot of an example entangling surface $\Sigma_1$, described by the function $R_{\text{UV}}(\theta)$ in polar coordinates, in the CGLP theory. {\bf (b)}  The extrinsic curvature $\kappa_{\text{UV}} (\theta)$ for the entangling surface $R_{\text{UV}}(\theta)$.  The extrinsic curvature is small over the whole curve.}
\label{curvProf}
\end{figure}

 As an example, we consider the entangling surface $\Sigma_1$ plotted in Figure~\ref{curvProf}, which has a small extrinsic curvature along the entire curve.
A good way of measuring the accuracy of the analytic approximation in \eqref{asympIntRth} is to compare the function $R_{\text{UV}}(\theta) - R_{\text{IR}}(\theta)$
\begin{figure}[htb]
\begin{center}
\leavevmode
\scalebox{1}{\includegraphics{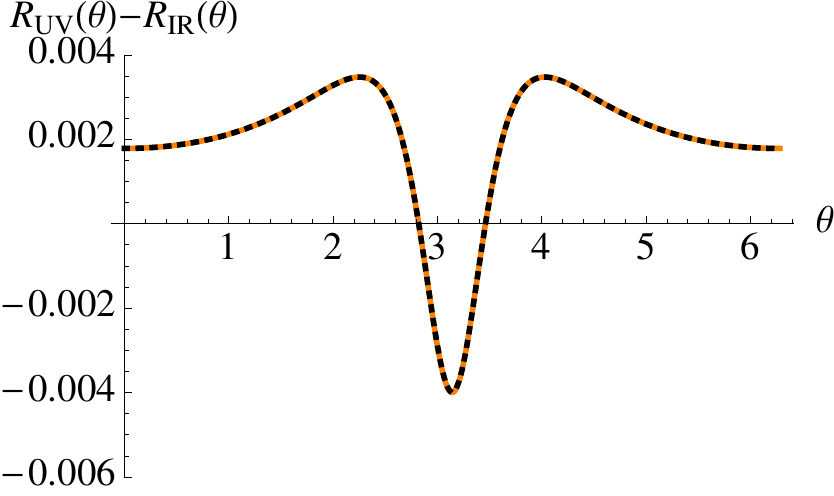}}
\end{center}
\caption{The function $R_{\text{UV}}(\theta) - R_{\text{IR}}(\theta)$ in the CGLP theory, with $ R_{\text{UV}}(\theta)$ plotted in fig.~\ref{curvProf}.  The solid orange curve is computed by numerically solving the equation of motion for the holographic entangling surface $\Sigma_2$.  The dotted black curve is an analytic approximation, which is equal to $ - \delta(\tau = 0, \theta)$, with $\delta(0,\theta)$ given in \eqref{asympIntRth}. }
\label{UVminusIR}
\end{figure}
 as computed both numerically and using \eqref{asympIntRth}, which is done in Figure~\ref{UVminusIR}.
Here the function $ R_{\text{IR}}(\theta)$ is the profile of the cylinder when it reaches $\tau = 0$: $ R_{\text{IR}}(\theta) = r(\tau = 0,\theta)$.  The analytic approximation simply gives $R_{\text{UV}}(\theta) - R_{\text{IR}}(\theta) \approx - \delta(\tau = 0, \theta)$, with $\delta(0,\theta)$ given in \eqref{asympIntRth}.  The two curves match extremely well.

\section{Shape dependence of the entanglement entropy in $(3+1)$-dimensional CFT}

The entanglement entropy for a smooth entangling surface $\Sigma$ in a $(3+1)$-dimensional CFT is given in \eqref{threegen}.
However, if the surface is not smooth, for example if it has conical or wedge singularities, then there may be additional contributions to \eqref{threegen}.  In this section we consider the entanglement entropy for a wedge and a cone in a $(3+1)$-dimensional CFT through both field theoretical and holographic computations.  We find that wedge entanglement entropy acquires a $1/\epsilon$ divergence not present in
\eqref{threegen}. The cone entanglement entropy has a $\log^2 \epsilon$ divergence as predicted by  \eqref{threegen}, but
its correct coefficient is twice smaller than for a regulated version of  \eqref{threegen}.

\subsection{The entanglement entropy for a wedge}\label{WedgeFT}

The wedge is the surface in $(3+1)$-dimensions given by
$(x^1,x^2,x^3) = (r\sin\phi, r\cos\phi,z)$, where $0\le r < \infty$, $\phi=0$ and $\Omega$, and $z \sim z  + L$.  We have compactified the $z$ direction on a large circle of length $L$ to avoid unwanted infrared divergences.  We begin by using the replica trick to calculate the entanglement entropy with this geometry for a free scalar field, and this is followed by a holographic computation.

\subsubsection{The free scalar field}

Using the replica trick one can show that the entanglement entropy for the massive scalar field is given by~\cite{Casini:2005rm}
\es{EEgen}{
	S =  \frac{1}{1-\alpha} \sum_{k=0}^{\alpha-1} \log Z_k \Big|_{\alpha\to1} \ ,
}
where $Z_k$ is the partition function of a scalar field $\phi_k$ on $4$-dimensional Euclidean space with boundary conditions
\es{bcscalar}{
	\phi_k (\vec x, t=0^+) = e^{2\pi i \frac{k}{\alpha}} \phi_k (\vec x, t=0^-) \ , \qquad \vec x \in A \ ,
}
where $A$ is the region bounded by the wedge.
Since the theory we consider is free, the partition function $Z_k$ is obtained from
the integral of the Green's function:
\es{ZFormula}{
	\frac{\partial}{\partial m^2} \log Z_k = - \frac{1}{2} \int d^{d+1} X \,G_k (\vec X, \vec X) \ .
}
The Green's function is the two-point correlation function of the free massive scalar
and is subject to the following conditions:
\es{Green}{
	(-\Delta_{\vec X} + m^2 )\, G_k (\vec X, \vec X') &= \delta (\vec X - \vec X') \ , \\
	\lim_{\epsilon \to 0^+} G_k( (\vec x, \epsilon), \vec X') &= e^{2\pi i \frac{k}{\alpha}}\lim_{\epsilon \to 0^-} G_k( (\vec X, \epsilon), \vec X') \ , \qquad \vec x \in A \ .
}

We expand the Green's function in Fourier modes along the $z$-direction.  The problem of finding the $4$-dimensional Green's function then reduces to that of finding the $3$-dimensional Green's functions for a cusp entangling surface for a tower of massive fields, with masses
\es{kkmasses}{
M_n^2 = m^2 + \left( { 2 \pi n \over L} \right)^2 \,, \qquad n\in \mathbb{Z} \,.
}
Using the result in~\cite{Casini:2006hu} for the Green's function with a cusp entangling surface in $3$-dimensions, we find
\es{GreenWedge}{
	G_k (\vec X, \vec X') = \frac{2}{L} \sum_\nu \sum_{n=-\infty}^\infty \int_0^\infty d \lambda \frac{\lambda}{\lambda^2 + M_n^2}  g_n(\vec x) g_n^\ast (\vec x)\ ,
}
where we use $\vec x$ to denote the $3$-dimensional coordinates $(t,x^1,x^2)$.
The $g_n$ are the eigenfunctions of the $3$-dimensional Laplace operator $(-\Delta_{\vec x} + M_n^2)$, whose eigenvalues we denote by $(\lambda^2 + M_n^2)$.
Using spherical coordinates with $(t,x_1,x_2)=(\rho\cos\theta, \rho\sin\theta\sin\phi, \rho\sin\theta\cos\phi)$, the eigenfunctions are given explicitly by
\es{gn}{
	g_n(\vec x) = \psi_\nu (\theta, \phi) \frac{J_{\frac{1}{2} + \nu} (\lambda \rho)}{\sqrt{\rho}} \ ,
}
where $J$ is the Bessel function of the first kind.
The functions $\psi_\nu$ are the eigenfunctions of the angular laplacian $\Delta_\Omega$ on the two-sphere,
\es{phinu}{
	\Delta_\Omega \psi_\nu (\theta,\phi) = - \nu (\nu+1) \psi_\nu (\theta,\phi) \,, \qquad \int  d\theta \, d\phi\, \sin(\theta) |\psi_\nu(\theta,\phi)|^2 = 1\,,
}
subject to the boundary condition
\es{phibc}{
	\lim_{\epsilon\to 0^+} \psi_\nu (\frac{\pi}{2} + \epsilon, \phi) = e^{2\pi i \frac{k}{\alpha}} \lim_{\epsilon\to 0^+} \psi_\nu (\frac{\pi}{2} - \epsilon, \phi) \ ,\qquad \phi \in [0,\Omega ] \ .
}

Preforming the integral over $\lambda$ in \eqref{GreenWedge} and taking a derivative of the partition function in \eqref{ZFormula} with respect to $m^2$ gives~\cite{Calabrese:2004eu}
\es{dZ}{
	\frac{\partial }{\partial m^2} \log Z_k = - \frac{L}{4 m}\coth \left( \frac{m L}{2} \right)  \sum_\nu \left( \nu + \frac{1}{2} \right) \ .
}
The sum over the eigenvalues $\nu$ is divergent and needs regularization.
The computation of the sum was carried out in detail in \cite{Casini:2006hu}, and
one finds that the regularized sum only depends on $k$ and the angle of the cusp $\Omega$.
After integrating \eqref{dZ} with respect to $m^2$, we find that
the entanglement entropy for a wedge has the angle dependent UV divergence
\es{wedge4d}{
	S_\text{wedge} =  \int^{1/\epsilon^2}dm^2 \frac{1}{\alpha-1} \sum_{k=0}^{\alpha-1}\left( \frac{\partial }{\partial m^2} \log Z_k\right)  \Big|_{\alpha \to 1} = f^\text{(scalar)}_\text{cusp}(\Omega)\frac{L}{\epsilon} + O(\epsilon^0)\ ,
}
where the function $f^\text{(scalar)}_\text{cusp}(\Omega)$ is the same function as for the cusp in $(2+1)$-dimensional CFT \cite{Casini:2006hu}.
It behaves as $f^\text{(scalar)}_\text{cusp}(\Omega) \sim 1/\Omega$ when the angle is
very small, while it becomes zero at $\Omega = \pi$ where there is no cusp in the entangling surface.
The function $f_\text{cusp}(\Omega)$ is not universal and depends on the type of matter.

\subsubsection{A holographic computation}
Next we compute the holographic entanglement entropy for the wedge.
To this end, we use the following $AdS_5$ metric,
\es{ads5}{
	ds^2 = \frac{dy^2 - dt^2 + dr^2 + r^2 d\phi^2 + dz^2}{y^2} \ .
}
For simplicity, we have set the AdS radius to 1. The central charges $a$ and $c$ of the dual CFT$_4$,
normalized as in \eqref{threegen}, are then determined by the 5-dimensional Newton constant \cite{Henningson:1998gx}.
\be
\label{centralcharges}
a = 3 \, c = \frac {45 \pi} {G_N^{(5)} }
\ .
\ee

The wedge is defined by $\Sigma = \{ 0\le r < r_{\text{max}},\, \phi = \pm\frac{\Omega}{2},\, z \sim z + L \}$ at the AdS boundary $y=0$.
 The large radius cut-off $r_{\text{max}}$ and the length $L$ are introduced to regularize the volume of the wedge. As usual, to introduce a UV cut-off we will restrict $y \geq \epsilon$.
The entanglement entropy functional is given by
\es{areawed}{
	S = {L \over 4G_{N}^{(5)}} \int dr \int d\phi \,  \frac{1}{y^3(r,\phi)}  \sqrt{ r^2 + r^2 (\partial_r y)^2 + (\partial_\phi y)^2} \ ,
}
where we take the holographic coordinate $y$ to be a function of $(r,\phi)$.

We must find the function $y(r,\phi)$ which minimizes the entanglement entropy functional and approaches the wedge at the boundary of $AdS_5$.
The scaling symmetry of the spacetime and the wedge implies the following ansatz for the minimal surface \cite{Drukker:1999zq,Hirata:2006jx}:
\es{ansatz}{
	y(r,\phi) = \frac{r}{ g(\phi )} \ .
}
With this ansatz the initial value problem becomes first order\footnote{Note that one must first find the equation of motion for $y(r,\phi)$ and then subsequently substitute the ansatz in \eqref{ansatz}.}
\es{eomwedge}{
	\frac{dg}{d\phi} = g\sqrt{(1+g^2)\left( \frac{g^2(1+g^2)^2}{g_0^2(1+g_0^2)^2} - 1 \right)} \,, \qquad g_0 = g(0) \,, \quad g'(0)=0 \,.
}
 It follows that the angle of the wedge determines $g_0$ as
\es{omg0}{
	\frac{\Omega}{2} &= \int_{g_0}^\infty dg \frac{1}{g\sqrt{(1+g^2)\left( \frac{g^2(1+g^2)^2}{g_0^2(1+g_0^2)^2} - 1 \right)}} \ .
}
Integrating this equation we find that, as in the $(2+1)$-dimensional cusp calculation \cite{Drukker:1999zq,Hirata:2006jx}, the limiting value where $g_0=0$ is $\Omega=\pi$.

The entanglement entropy is then found by evaluating the regularized functional in \eqref{areawed} on the solution to the equation of motion:
\es{areawed2}{
	S  &= \frac{2 L}{4G_N^{(5)}} \int_{g_0\epsilon}^{r_{\text{max}}} \frac{dr}{r^2} \int_{g_0}^{r/\epsilon} dg\,  h(g,g_0) \\
	&= \frac{2 L}{4G_N^{(5)} \epsilon} \int_{g_0}^{r_{\text{max}}/\epsilon} \frac{dr}{r^2} \left[ \frac{r^2 - g_0^2}{2} +  \int_{g_0}^{r} dg\,   (h(g,g_0) - g) \right]\\
	&=  \frac{1}{4G_N^{(5)}} \left[ \frac{A_\Sigma}{2\epsilon^2} -  f^\text{(hol)}_\text{wedge}(\Omega) \frac{L}{\epsilon} + O(\epsilon^0)\right]\ .
}
The functions $h(g,g_0)$ and $f^\text{(hol)}_\text{wedge}(\Omega)$ are defined by
\es{fom}{
	h(g,g_0)&= \frac{g^2 (1+g^2)}{ \sqrt{g^2(1+g^2)^2 - g_0^2(1+g_0^2)^2}} \ ,\\
	f^\text{(hol)}_\text{wedge}(\Omega) &= g_0 - \int_{g_0}^\infty \frac{dr}{r^2} \int_{g_0}^{r} dg\,  (h(g,g_0) - g) \ .
}

\begin{figure}[htb]
\begin{center}
\leavevmode
\scalebox{0.9}{\includegraphics{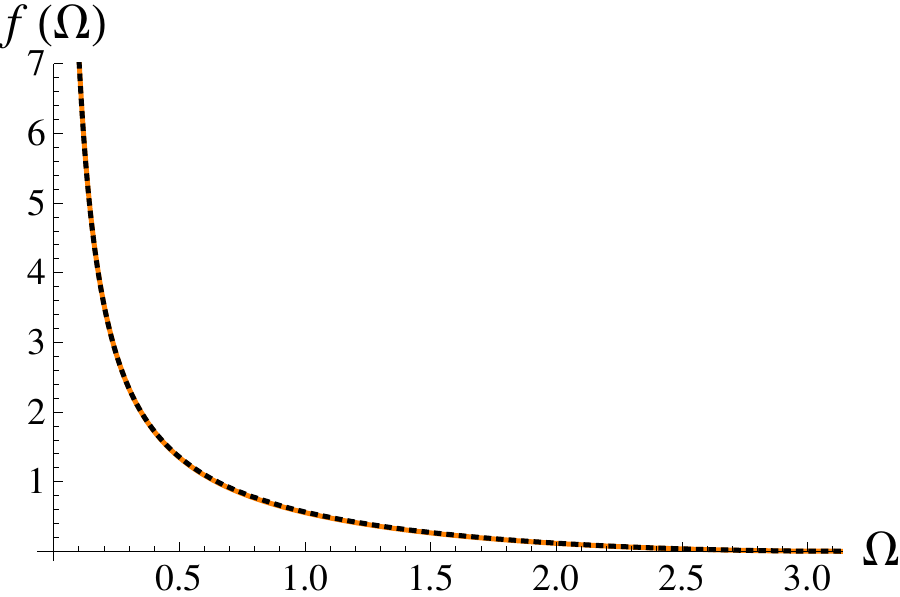}}
\end{center}
\caption{The entanglement entropy for the wedge has a $1/\epsilon$ divergent term, whose coefficient depends on the angle of the wedge. This coefficient $f_\text{wedge}^\text{(hol)}(\Omega)$ is given explicitly in \eqref{omg0} and \eqref{fom}, and its normalization depends on the UV cut-off $\epsilon$.
The black dotted line is the normalized function $\tilde f_\text{wedge}^\text{(hol)}(\Omega)= a f_\text{wedge}^\text{(hol)}(\Omega)$ for the wedge with $a\sim1.11$ and the orange line is the function $f_\text{cusp}^\text{(hol)}(\Omega)$
for the cusp in $(2+1)$-dimensions.
}
\label{Wedge4d}
\end{figure}

Just like the free scalar field calculation for the wedge, \eqref{wedge4d}, the holographic result has an
$L/\epsilon$ divergence that was absent for smooth entangling surfaces.\footnote{
Since on the sides of the wedge the extrinsic curvature vanishes, it is reasonable to think of this term in the entanglement
entropy as due to the wedge singularity.}
A numerical plot of the function $f^\text{(hol)}_\text{wedge}(\Omega)$ is shown in Figure~\ref{Wedge4d}, where it can be seen that
it goes to zero as $\Omega$ approaches $\pi$.
When $\Omega$ is small
$f^\text{(hol)}_\text{wedge}(\Omega)$ diverges as
\es{fsmall}{
	f^\text{(hol)}_\text{wedge}(\Omega) \sim \frac{0.646}{\Omega} \ .
}
A pole at $\Omega = 0$ also appeared in the field theory computation \eqref{wedge4d}.  With that said, the function $f^\text{(hol)}_\text{wedge}(\Omega)$ is different from that of the scalar field theory, $f_\text{cusp}^\text{(scalar)}(\Omega)$, which appeared in \eqref{wedge4d}.

 A surprising result, however, is that after an overall rescaling the function $f^\text{(hol)}_\text{wedge}(\Omega)$ agrees with function $f^\text{(hol)}_\text{cusp}(\Omega)$ in \cite{Hirata:2006jx} describing the holographic cusp anomaly in $(2+1)$-dimensions. The normalization of the function $f_\text{wedge}(\Omega)$ depends on the choice of the UV cut-off scale $\epsilon$. We introduce the normalized function $\tilde f_\text{wedge}^\text{(hol)}(\Omega) = a f_\text{wedge}^\text{(hol)}(\Omega)$ by tuning the constant $a$ such that $\tilde f_\text{wedge}^\text{(hol)}(\Omega)$ agrees with $f_\text{cusp}^\text{(hol)}(\Omega)$ in the limit of $\Omega \to 0$. We find $a\sim 1.11$ numerically, and in Figure~\ref{Wedge4d} we plot both $\tilde f_\text{wedge}^\text{(hol)}(\Omega)$ and $f_\text{cusp}^\text{(hol)}(\Omega)$.
The plot shows that in fact the normalized function $\tilde f_\text{wedge}^\text{(hol)}(\Omega)$ is the same 
(within the numerical accuracy) as $f_\text{cusp}^\text{(hol)}(\Omega)$, although the definitions
\eqref{omg0} and \eqref{fom} appear quite different from those for the cusp in \cite{Hirata:2006jx}.
 For a free scalar field the function $f_\text{wedge}^\text{(scalar)}(\Omega)$ for the wedge also turned out to be the same as  $f_\text{cusp}^\text{(scalar)}(\Omega)$ for the cusp (see \eqref{wedge4d}). It is very interesting that the appropriately normalized functions $f(\Omega)$ agree for the cusp and wedge geometries both in the free and in the strongly coupled theories that we have studied.

\subsection{The entanglement entropy for a cone}\label{ConeFT}

In this section we show that when the entangling surface in $(3+1)$-dimensional CFT has a conical singularity, the entanglement entropy acquires a $\log^2 (r_{\text{max}}/\epsilon)$ divergence.
We take the entangling surface to be the cone defined by $(r,\theta,\phi) = (r, \Omega ,\phi)$, where $0\le r < r_{\text{max}}$, $\phi\sim \phi + 2\pi$ is the azimuthal angle, and $\Omega$ is the opening angle of the cone.  The large distance cut-off $r_{\text{max}}$ regulates the area of the cone.

To begin, we will evaluate \eqref{threegen} for this surface.  Even though this equation is only valid for smooth entangling surfaces, it does provide a quick way of seeing how the $\log^2\epsilon$ divergence appears.  The cone has two normal vectors in $\mathbb{R}^{1,3}$, given by $n_\mu^1 = (1,0,0,0)$ and $n_\mu^2 = (0,0,r,0)$ in $(t,r,\theta,\phi)$ coordinates.  Only the second fundamental form associated with $n_{\mu}^2$ is non-vanishing, with non-zero component $k^2_{\phi \phi} =  \frac{1}{2}r\sin 2\Omega$.
 The $c$-anomaly term in \eqref{threegen} then contributes
\es{conelog2}{
&\frac{c}{480\pi}\log \epsilon \int_{r_0}^{r_{\text{max}}} dr \int_0^{2\pi} d\phi \, r\sin \theta (k^2_{\phi\phi})^2 (g^{\phi\phi})^2 \\
	& =  -\frac{c}{240} \frac{\cos^2\Omega}{\sin\Omega}
\log^2 \epsilon
+ \cdots \ .
}
In going from the first to the second line in \eqref{conelog2}, we assume that the tip of the cone is cut-off at some short distance $r_0 \propto \epsilon$.  Note that the $a$-anomaly does not give an additional contribution to the singularity because $\int R_\Sigma = 0$.
The new UV divergent term \eqref{conelog2} vanishes at $\Omega = \frac{\pi}{2}$, where there is no conical singularity, while its coefficient diverges as $\Omega$ goes to zero.

A more heuristic way to obtain the $\log^2 \epsilon$ term, similar to an argument for the cusp geometry in \cite{Casini:2006hu} is as follows.
When $\Omega$ is small, the cone may be approximately decomposed into a union of cylinders with radius $R=L \sin\Omega$ and
the length $\Delta L$, where $L$ is the length from the apex of the cone to one of the cylinders and
$\Delta L$ is supposed to be small.
The logarithmic term of the entanglement entropy of the cylinder comes from the $c$-anomaly given in \eqref{threegen}
\es{logCyl}{
	S_{\text{cylinder}} = \frac{c}{240} \frac{\Delta L}{R}\log(R/\epsilon) \ .
}
It follows that the entanglement entropy of the cone has the square of the logarithmic divergence
\es{log2}{
	S_{\text{cone}} \approx \int_\epsilon^{r_\text{max}} dL\, \frac{c}{240} \frac{1}{L\sin\Omega}\log(L\sin\Omega/\epsilon) =  - \frac{c}{480} \frac{1}{\sin\Omega}
\log^2 (r_\text{max} /\epsilon)
+ \cdots \ ,
}
which reproduces the leading behavior of \eqref{conelog2} in the small $\Omega$ limit, except it is smaller by a factor of 2.
We will see below that the factor in \eqref{log2} is correct.

We now present a more precise holographic derivation of the $\log^2\epsilon$ divergence, which correctly takes into account the conical singularity.
 We use the $AdS_5$ metric
\es{ads5cone}{
	ds^2 = \frac{dy^2 - dt^2 + dr^2 + r^2 (d\theta^2 + \sin^2\theta d\phi)}{y^2} \,,
}
so that the entangling surface is given by the cone $\Sigma = \{ 0\le r < \infty,\, \theta=\Omega,\, 0\le \phi < 2\pi \}$ at the AdS boundary $y=0$.
Taking the holographic coordinate $y$ to be a function of $(r,\theta)$,
we use the conformal and rotational symmetries of AdS spacetime and the entangling surface to restrict the ansatz to
\es{ansatzcone}{
	y(r,\theta) = \frac{r}{\tilde g(\theta)} \ .
}
The entanglement entropy functional is then given by
\es{areacone}{
	S &= {2\pi \over 4 G_N^{(5)} }\int \frac{dr}{r} \int d\theta \sin\theta\, \tilde g \sqrt{\tilde g^4 + \tilde g^2 + (\tilde g')^2} \\
	&= -{\pi \over 2 G_N^{(5)} } \int \frac{dr}{r} \int ds\, g \sqrt{g^4 + g^2 + (1-s^2)(g')^2}\\
	&= -{\pi \over 2 G_N^{(5)} }   \int \frac{dr}{r} \int dg\, g\sqrt{(g^4 + g^2)s'(g)^2 + 1 - s(g)^2} \ ,
}
where we introduced the new variable $s = \cos \theta$, which runs from $s_0 \equiv \cos\Omega$ to unity,
and redefined $g(s) = \tilde g(\theta(s))$.
In the last equality, we changed the integration variable from $s$ to $g$.

The entanglement entropy is given by evaluating the EE functional in the last line of~\eqref{areacone} on the function $s(g)$ which solves the Euler-Lagrange equation,
\es{ELcone}{
{g \,s(g) \over \sqrt{(g^4 + g^2)s'(g)^2 + 1 - s(g)^2} } + \frac{d}{dg} \left[ {g (g^4 + g^2) s'(g) \over \sqrt{(g^4 + g^2)s'(g)^2 + 1 - s(g)^2} } \right] = 0\,,
}
subject to the boundary condition $s(g=r/\epsilon) = s_0$, where the UV cutoff is put at $y=\epsilon$ in the AdS spacetime.  However, to find the leading divergences in the entanglement entropy, it only is necessary to know the function $s(g)$ near the boundary at $g =r/\epsilon $, which we may assume is a large number.  Taking the large $g$ limit of~\eqref{ELcone}, one may verify that the asymptotic expansion for $s(g)$ near the boundary is
\es{gexp}{
	s(g) =  s_0 + \frac{s_0}{4 g^2} + O \left(\frac{\log g}{g^4}\right) \ .
}
While the solution above only satisfies the boundary conditions up to a term of order $(\epsilon / r)^2$, the difference does not affect the leading two singular terms in the EE.
In evaluating the entanglement entropy functional in \eqref{areacone} on the solution $s(g)$ in \eqref{gexp}, we must evaluate the integral
\es{sint}{
	\int_{r/\epsilon}^{g_0}  dg \, \left[ \sqrt{1-s_0^2}\, g - \frac{s_0^2}{8\sqrt{1-s_0^2}} \frac{1}{g} +
	O \left( \frac{\log g}{g^3} \right) \right]
	= - \frac{\sin\Omega}{2} \frac{r^2}{\epsilon^2} +  \frac{\cos^2\Omega}{8\sin\Omega} \log (r/\epsilon) +O(\epsilon^0) \ ,
}
where $g_0$ is the minimum value of $g(s)$.
Then, preforming the $r$ integral from $r=g_0 \epsilon$ to
$r_{\text{max}}$, we obtain the entanglement entropy for
the cone:
\es{areaconeUV}{
	S_{\text{cone}} = \frac{1}{4G_N^{(5)}} \left[ \frac{A_\Sigma}{2\epsilon^2} - \frac{\pi \cos^2\Omega}{8\sin\Omega}\log^2 (r_{\text{max}}/\epsilon) + \ldots
\right]\ .
}

In order to compare this result with the naive calculation in \eqref{conelog2}, we use \eqref{centralcharges}.
One then sees that \eqref{areaconeUV} is smaller than \eqref{conelog2} by a factor of 2.
As stressed above, the approach of \eqref{threegen} is not precise for singular entangling surfaces. It is nice, therefore, that it is
only a factor of 2 off from the precise holographic result \eqref{areaconeUV}.

\section*{Acknowledgments}
We thank H.~Liu, J.~Maldacena, M.~Mezei, R.~Myers, and E.~Witten
 for helpful discussions, and especially H.~Casini for very useful comments on this manuscript.
 The work of IRK and TN was supported in part by the US NSF under Grant No.~PHY-0756966. IRK gratefully acknowledges support from the IBM Einstein Fellowship at the Institute for Advanced Study, and from the John Simon Guggenheim Memorial Fellowship.  SSP was supported by a Pappalardo Fellowship in Physics at MIT and by the U.S. Department of Energy under cooperative research agreement Contract Number DE-FG02-05ER41360\@.  BRS was supported by the NSF Graduate Research Fellowship Program. BRS thanks the Institute for Advanced Study for hospitality.

\appendix

 \section{Perturbed CFT} \label{confsec}

 In this appendix we discuss the entanglement entropy of $(2+1)$-dimensional CFTs which have gravitational duals, perturbed by relevant operators.  Many of these results were recently found in~\cite{Liu:2012ee}.

We work with an effective $(3+1)$-dimensional gravitational theory, with metric
\es{D4Metric2}{
ds_{4}^2 &= \frac{L^2_{\text{UV}}}{y^2} \left( \frac{dy^2}{f(y)} -dt^2 + dr^2 + r^2 d\theta^2 \right) \,, \qquad y \geq 0 \,, \\
}
and we assume that in the UV (small $y$) the metric asymptotes to $AdS_4$ with radius $L_{\text{UV}}$, i.e.
$f(y) = 1 + O(y^\alpha)$, $\alpha > 0$.
The entanglement entropy across a circle of radius $R$ in the boundary QFT is then given by the area functional
\es{areaGEN}{
S(R) = {\pi L_{\text{UV}}^2 \over 2 G_N^{(4)}} \int_{\epsilon}^{y_{{\text{IR}}}} d y \,{ r(y) \over y^2 \sqrt{f(y)}} \sqrt{1 + f(y) (\partial_y r)^2}\,,
}
where the function $r(y)$ satisfies the Euler-Lagrange equation, $\epsilon$ is the short-distance cut-off, and $y_{\text{IR}}$ is the maximal value of $y$ for the solution.

If the effective $(3+1)$-dimensional gravitational theory comes from an exact $D$-dimensional string or M-theory background, with metric as in \eqref{ESframe}, then we may identify
\es{fyrelation}{
 f(y) = \beta(u) \left( { \partial y \over \partial u} \right)^2 \,, \qquad  {L_{\text{UV}}^2 \over y^2} =  \frac{V(u)} {V_{\text{UV}}}    \alpha(u) \beta(u) \,.
}

Let us begin with the conformal limit, where we may take $f(y) = 1$. We also define the $(3+1)$-dimensional Newton's constant $G_N^{(4)} = G_N^{(D)} / V_{\text{UV}}$, where, in $10$- and $11$-dimensions, the Newton's constant takes the values $G_N^{(10)} = 8 \pi^6 \alpha'^4 g_s^2$ and $G_N^{(11)} = (32 \pi^2)^{-1} (2 \pi \ell_p)^9$, respectively.
The solution to the equation of motion is then $r(y) = \sqrt{ R^2 - y^2}$.  Evaluating the area functional on this solution and expanding in $\epsilon$ gives
\es{SRconf}{
S(R) = {\pi L_{\text{UV}}^2 \over 2 G_N^{(4)}} R  \int_\epsilon^R  \frac{dy}{y^2} =  {\pi R L^2_{\text{UV}} \over 2 G_N^{(4)}\epsilon} -{\pi L_{\text{UV}}^2 \over 2 G_N^{(4)}}\,.
}

Suppose that the 3-dimensional CFT comes from the near horizon limit of $N$ M2-branes at the tip of the cone $\mathcal{C}_Y = \mathbb{R} \times Y$, where $Y$ is some 7-dimensional internal Sasaki-Einstein space.  In the large $N$ limit the theory is well described by the supergravity background
\es{sugra}{
ds_{11}^2 = ds_{AdS_4}^2 + 4 L_{\text{UV}}^2 ds_Y^2 \,, \qquad F_4 = \frac{3}{L_{\text{UV}}} \vol_{AdS_4} \,, \qquad F_7 = *_{11} F_4 = 384 L_{\text{UV}}^6 \vol_Y \,,
}
where the radius $L_{\text{UV}}$ of $AdS_4$ is quantized in plank units:
\es{quant}{
N = \frac{1}{(2 \pi \ell_p)^6} \int_Y F_7 = \frac{6 \Vol(Y)}{\pi^6} \frac{L_{\text{UV}}^6}{\ell_p^6} \,.
}
Substituting the relation between $N$ and $L_{\text{UV}}$ in~\eqref{quant} into~\eqref{SRconf} gives the renormalized entanglement entropy
\es{SRconf1}{
F_{\text{UV}} = {\pi L_{\text{UV}}^2 \over 2 G_N^{(4)}} = N^{3/2} \sqrt{ \frac{2 \pi^6}{27 \Vol(Y)}} + O(N^{1/2}) \,.
}
As a consistency check, we may verify that
this is equal to the finite part of the Euclidean free energy of the theory on the $3$-sphere \cite{Herzog:2010hf, Jafferis:2011zi},
in agreement with the general result of \cite{Casini:2011kv}.

Now suppose that there is an RG flow in the boundary QFT  caused by perturbing the UV CFT by a relevant scalar operator ${\cal O}$ of dimension $1/2 \leq \Delta < 3$.  The operator ${\cal O}$ is dual to a massive scalar field in the bulk, with $\Delta (\Delta - 3) = m^2$.  The bulk action is then given by
\es{bulkaction}{
I_4 = \frac{1}{16\pi G_N^{(4)}}\int d^{4}x\, \sqrt{-g} \left[R + \frac{6}{L_{\text{UV}}^2} - \frac12 g^{\mu \nu} \partial_\mu \phi \partial_\nu \phi  - \frac12 m^2 \phi^2 + \cdots \right] \,,
}
where the dots stand for higher order terms in $\phi$ and the contributions of other fields, which won't be relevant for this discussion.  When the field $\phi$ vanishes the equation of motion for the metric simply gives $AdS_4$.  The equations of motion for $\phi$ in the $AdS_4$ background gives the following asymptotic solution at small $y$,
\es{solnPhi}{
\phi(y,x) = y^{3-\Delta} [ \phi_0(\vec x) + O(y^2) ] \,,
}
where we have chosen the solution which stays finite at the boundary.  This is the solution which corresponds to perturbing the action of the UV boundary CFT to
\es{boundaryPert}{
I_3  = I_{\text{UV}}+ \int d^3x \phi_0(\vec x) {\cal O}(\vec x) \,.
}
The field $\phi$ has a back-reaction on the metric.  In particular, when $\phi_0(\vec x) = \phi_0$ is constant, we find that at small $y$ the Einstein equation gives us
\es{fexpand}{
f(y) = 1 + \frac{3-\Delta}{4 } \phi_0^2 \, y^{2 (3 - \Delta)} + \cdots \,.
}

To find the first correction to the renormalized entanglement entropy as a result of the relevant deformation of the UV CFT, it is sufficient to evaluate the action in \eqref{areaGEN} with $f(y)$ given in \eqref{fexpand} on the UV solution $r(y) = \sqrt{R^2 - y^2}$.  A straightforward calculation then gives
\es{FFirstOrder}{
{\cal F}(R) = F_{\text{UV}} \left( 1 - \frac{(3-\Delta) }{8 \left( \frac72  - \Delta\right) } \phi_0^2 \, R^{2(3-\Delta)} + \cdots \right) \,.
}
The factor $3-\Delta$ plays an important role; it ensures that the renormalized entanglement entropy is not changed by marginal
perturbations.

\bibliographystyle{ssg}
\bibliography{CGLP}

\end{document}